\documentclass[11pt, oneside]{article}   	
\usepackage{geometry}                		
\usepackage{graphicx}				
\usepackage{amssymb}
\usepackage{amsmath}

\title{A possible mechanism of the lifetime effects of electron pairs at the SrTiO\textsubscript{3}/LaTiO\textsubscript{3} interface}
\author{Xing Yang}
\date{\today}							

\begin{document}
\maketitle
\abstract{
The lifetime effects of the electron pairs are supposed to be responsible for the atypical behavior of the single electron transistors manufactured at the strontium titanate/ lanthanum aluminate (STO/LAO) interface. In addition, as shown in Richter's experiments, the energy gap $\Delta$ should persist above the superconducting transition temperature $T_c$. In order to explain the experiments, the present paper attempts to associate the phonon-electron interactions to the mechanism of the formation and decay of the electron pairs. Moreover, through employing the boson fermion model that includes two superconducting mechanisms, we reproduce the step-like $\Delta-T_c$ relations with the energy gap $\Delta$ remaining finite above $T_c$. The results are roughly consistent with the experiments. The newly-introduced models and formalism may be helpful in describing superconductors with lifetime effects and a finite energy gap $\Delta$ above $T_c$. }

\section{introduction}
Immediately after the Bardeen-Cooper-Schrieffer (BCS) theory was established \cite{BCS}, the number parity effects were expected in superconductors. Basically, the BCS ground state corresponds to a coherent of superposition of pair states with even number parity, while the total number of quasi-particles $N$ is not fixed. Furthermore, if the BCS ground state is projected into odd and even parity states \cite{Schrieffer}, one has to differentiate two cases: (a) the total number of quasi-particles $N$ is even. (b) $N$ is odd. Remarkably, the electron pairing in superconductors results in great differences between the two cases. In the first case, all electrons are paired and form the electron pair states, while in the second case, one unpaired electron is left and the ground state will include one more Bogoliubov quasi-particle except the electron pairs.

Not surprisingly, the difference between even and odd fermionic states is first discovered in nuclear physics by Bohr, Mottelson and Pines \cite{Bohr:1958aa}, and its total number of particles $N$ is around $10^2$. Subsequently, with $N\sim 10^9$, the number parity effects are observed in the variation of the tunneling current through one Coulomb-blockaded mesoscopic superconducting island in single electron transistors (SETs) by Mooij et al. \cite{Geerligs:1990aa}, Tinkham and his coworkers \cite{Tuominen:1992aa}, as well as Devoret and his colleagues \cite{Lafarge:1993ab,Lafarge:1993aa}. The measured even/odd free energy difference is found to decrease linearly with the temperature $\delta {\cal F}_{\rm e/ o} \sim \Delta_0 - k_B T \log N_{\rm eff}$. Here $\Delta_0$ is the energy gap at low temperature and $N_{\rm eff}$ is the number of effective excitation states for the unpaired electron in the odd parity states. From the formula, the even-odd effects will be observed below the crossover temperature $T^* \sim \Delta_0/(k_B \log N_{\rm eff})$ at which the even/odd free energy difference is vanishingly small. For a typical SET with an aluminum superconducting island \cite{Grabert1992}, the crossover temperature $T^*$ is approximately $10^2 {\rm mK}$ and the superconducting transition temperature is around $1 {\rm K}$. Consequently, it is estimated that $T^* \ll T_c$ and $N_{\rm eff}\sim10^4$\cite{Tuominen:1992aa,Lafarge:1993ab,Lafarge:1993aa,Janko:1994aa}.

On the contrary, Levy's experiments at the strontium titanate/lanthanum aluminate (STO/LAO) interface exhibit huge distinctions compared with the previous experiments \cite{Cheng:2015aa}. The crossover temperature $T^*$ is detected to be $\sim 900 {\rm mK}$, much higher than the superconducting transition temperature $T_c \sim 300{\rm mK}$. That is, the even-odd effects are observed without superconductivity, while the electron pairing is still preserved. Moreover, the extracted $N_{\rm eff}$ from experiments is $\sim 1$, much smaller than that of the typical SETs.

In addition, the superconducting gap $\Delta$ of the electron system at the STO/LAO interface are detected of the order of $40 {\rm \mu eV}$ and its V-shaped density of states implies the formation of pseudogap states \cite{Richter2013,Richter:2013aa}. Furthermore, the gap increases with charge carrier depletion in both the negatively and positively charged region. The experimental results are analogous to the behavior of the high-temperature superconductors. Beside that, the superconducting gap $\Delta$ remains finite in both electron systems at the STO/LAO interface and in high-$T_c$ superconductors.

The microscopic superconducting origin of the electron system at STO/LAO interface is still under discussions, while many theoretical proposals explain the experimental facts very well with self-consistency. Kedem, {\it et al.} interprets the mechanism as the coupling to the ferroelectric mode \cite{PhysRevLett.115.247002,PhysRevB.98.220505}. Simultaneously, Arce-Gamboa and Guzm\'an-Verri calculate the influence of the strain force to the ferroelectric mode and attained the phase diagram with the parameters of superconducting transition temperature and cation substitutions \cite{PhysRevMaterials.2.104804}. Conversely, Ruhman and Lee suggest a plasmon-induced superconducting mechanism \cite{PhysRevB.94.224515}. Additionally, W\"ofle and Balastsky connect the superconducting mechanism to transverse optical phonons \cite{PhysRevB.98.104505}.

Although the theoretical proposals are in good agreement with most experimental facts, some minor points are missing \cite{Yang}. Particularly, the energy gap should remain finite above superconducting transition temperature $T_c$. As we discussed, the finite gap above $T_c$ are crucial for the explanations of both the even-odd effects and scanning tunneling spectroscopy experiments. Also the von Hove singularity should be broadened in order to result in a small $N_{\rm eff}$ as suggested by Levy's experiments.

Continuing with the discussion of the reference \cite{Yang}, the paper clarified that the interaction potential in the phenomenological boson fermion model \cite{Alexandrov:1996} is induced by the phonon-electron interaction, while the microscopic origin of the primary superconducting gap $\Delta$ is still elusive. However, within the framework of perturbation theory, its relation with $T_c$ is obtained. The results are in good agreement with the experiments.

The paper is arranged in four sections. In Section \ref{sec:pip}, the phonon-induced interaction potential are deduced and its relation to even-odd effects discussed. In Section \ref{sec:tr}, the gap equation is calculated and the relation between the superconducting gap $\Delta$ and the transition temperature $T_c$ obtained and plotted. Finally, we summed up the content and draw our conclusions.

\section{Phonon-induced interaction potential}
\label{sec:pip}
\subsection{origin of phonon-induced interaction potential}
For the elelctron-phonon interaction, Fr\"ohlich has proposed a formula of the second-order correction to energy, in order to describe an electron emits and re-absorbed a phonon\cite{Bardeen:1951aa}\cite{Frohlich:1950aa}. The electron has momentum $\mathbf{k}$ and energy $\epsilon_\mathbf{k}$, which emits a phonon with momentum $\mathbf{q}$ and energy $\omega_{ph} = v_s |q|$ where the Planck constant $\hbar$ is set to be unity and $v_s$ is the speed of the sound. The intermediate state after the emission of the phonon has momentum

\begin{equation}
\mathbf{p}=\mathbf{k}-\mathbf{q}
\end{equation}

and its energy is denoted as $\epsilon_\mathbf{p}$. With the formula of the second-order perturbed energy

\begin{equation}
E=-2\sum_\mathbf{k}\sum_\mathbf{q}\frac{|M_{\mathbf{k},\mathbf{p}}|^2 P_\mathbf{k}P_\mathbf{p}}{\epsilon_\mathbf{p}-\epsilon_\mathbf{k}+\omega_{ph}}
\label{eq:1}
\end{equation}

where $M_{\mathbf{k},\mathbf{p}}=<\psi_\mathbf{p}|\hat{H}_I|\psi_\mathbf{k}>$ and $|\psi_\mathbf{k}>$ is the Bloch electron wave vector with momentum $\mathbf{k}$. The factor of two accounts for the two possible values of electron spin. $P_{\mathbf{p}}$ denotes the possibility to occupy a quantum state $|\psi_\mathbf{p}>$ and $P_{\mathbf{k}}$ is the average occupation number of a quantum state $|\psi_\mathbf{k}>$.

In Fr\"ohlich's case, the electron emits one phonon and is scattered into another quantum state, while in our case, the electron condenses into one boson after emitting a phonon. It results in the possibility to occupy a quantum state after condensing into a boson $P_{\mathbf{p}}=1$. Furthermore, if the former quantum state with momentum $\mathbf{k}$ is occupied, the average occupation number $P_{\mathbf{k}}=1$. Additionally, the coupling strength $M_{\mathbf{k},\mathbf{p}}$ can be put out of the summation sign:

\begin{equation}
E=-2|M|^2\sum_\mathbf{k}\sum_\mathbf{q}\frac{1}{\epsilon_\mathbf{p}-\epsilon_\mathbf{k}+\omega_{ph}}.
\label{eq:2}
\end{equation}

The energy of the intermediate state $\epsilon_\mathbf{p}$ is evaluated to be $\xi_0$ where $\xi_0$ is the ground state energy of the bosonic field. Note that the dispersion of the boson field is $\xi_\mathbf{t}=\xi_0+v |\mathbf{t}|-\mu_b$ where $v$ is the velocity of the bosonic condensates and $\mu_b$ is the chemical potential of the bosonic field. Eq. (\ref{eq:2}) becomes 

\begin{equation}
E=-2|M|^2\sum_\mathbf{k}\sum_\mathbf{q}\frac{1}{\xi_0-\epsilon_\mathbf{k}+\omega_{ph}}.
\end{equation}

Transforming the summation into integral over $\mathbf{q}$ and applying the dispersion of the phonons, it obtains:

\begin{equation}
E=\frac{L|M|^2}{\pi v_s}\sum_\mathbf{k}\ln(1-\frac{\epsilon_\mathbf{k}}{\xi_0}).
\end{equation}

Notice that the summation is over all quantum state of occupied electron state, and for a single electron with energy $\omega$, the phonon-induced interaction potential is

\begin{equation}
V_1=V_c \ln(1-\frac{\omega}{\xi_0})
\label{eq:3}
\end{equation}

where $V_c=\frac{L|M|^2}{\pi v_s}$. Notice that the upper limit of the integral is $\omega$, since one electron cannot emit a phonon with energy larger than the electron's energy. The lower limit of the integral is 0. The ground state energy of bosons $\xi_0$ is estimated to be $\sim 3\Delta$, the discussions of the part $\omega>\xi_0$ is meaningless, since at low temperature, the average occupation number $P_{\mathbf{k}}$ is vanishingly small. This leads to the cancellation of the second-order perturbed energy.

\subsection{even-odd effects with phonon-induced interaction potential}
\begin{figure}
  \includegraphics[width=240pt]{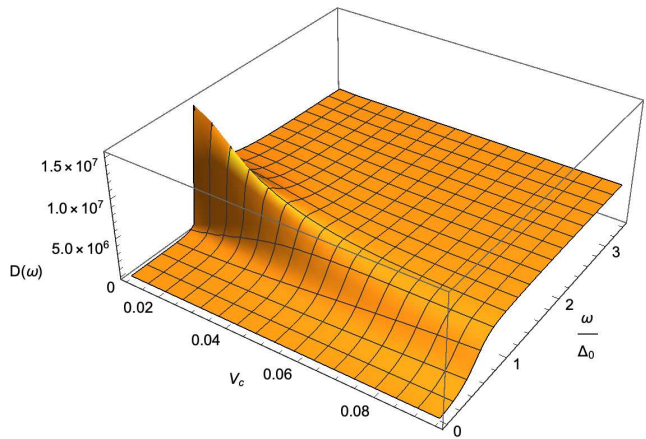}
  \caption{The density of states, $D(\omega)$, is plotted along with the variation of $V_c$. $D(\omega)$ and $V_c$ are plotted in atomic units.}
  \label{bfdos}
\end{figure}

The one-\cite{Yang} and two-dimensional \cite{josephson} phenomenological boson fermion model are presented with the formalism for computing the density of states in reference. With the replacement of the interaction potential from a phenomenological one to the phonon-induced potential, the results are plotted in Fig. \ref{bfdos}. Both the densities of state basically remain the same, including the broadening of the van Hove singularity and the production of the gap states.

\begin{figure}
  \includegraphics[width=250pt]{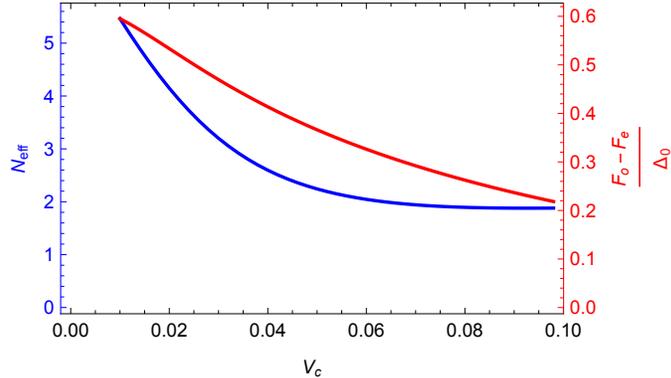}
  \caption{the even-odd free energy difference and the effective excitation number for the unpaired electron versus $V_c$. It is plotted in atomic units. $m^*= m_e$, $n_0=500$, $L=530 {\rm nm}$, $v_f=8.8 \times 10^3 {\rm m/s}$, $\Delta=40{\mu eV}$ and $v=0.073c$, where $c$ is the speed of light in vacuum.}
  \label{bf}
\end{figure}

Furthermore, the even-odd effects can be measured quantitatively with the even/odd free energy difference $\delta F_{e/o}$ and the number of excitation states for the unpaired electron in odd parity $N_{\rm eff}$. Both of the quantities can be determined by the density of states as shown in Ref. \cite{Yang}. In Fig. \ref{bf}, both of the quantities on even-odd effects are calculated  and plotted. The replacement of the interaction potential does not affect the conclusion in Ref. \cite{Yang}. That is, The lifetime effects result in the broadening of the van Hove singularity, and the broadening can produce a small $N_{\rm eff}$.

On the other hand, the interaction kernel can be caused by other interactions as well. If the kernel is zero at Fermi surface and finite when $\omega>0$, the even/odd free energy difference $\delta F_{e/o}$ and the effective excitation number $N_{\rm eff}$ can be easily generated and consistent with the experiments. It is one of the reasons that the present phonon-induced kernel may not be the only microscopic origin for the lifetime effects, while other interactions may also participate in the process.

\section{$\Delta-T_c$ relation}
\label{sec:tr}
\subsection{gap equation}
The starting point of the deductions is an equation that defines superconducting energy gap,

\begin{equation}
\label{eq:411}
\Delta_s \equiv \sum_k \frac{V_1(q=0)}{\sqrt{n_0}}F(k,\tau=0)
\end{equation}

where $F(k,\tau)=<\hat{c}_{-\mathbf{k}\downarrow}(0)\hat{c}_{\mathbf{k}\uparrow}(\tau)>$ is the anomaly Green's function and $\hat{c}_{\mathbf{k}\uparrow}(\tau)(\hat{c}^\dagger_{\mathbf{k}\uparrow}(\tau))$ denotes the annihilation (creation) operator of the quasiparticles in the boson-fermion model. The time derivative of the anomaly Green's function is

\begin{equation}
\label{eq:412}
\frac{\partial F(k,\tau)}{\partial \tau}=-E_p F-\frac{V_1^\dagger(q=0)}{\sqrt{n_0}}<\hat{b}_0>G
\end{equation}

where $G=<T_\tau[\hat{c}_{-\mathbf{k}\downarrow}(0)\hat{c}_{-\mathbf{k}\downarrow}(\tau)]>$ and $T_\tau[...]$ is the time-ordering operator. The operator $\hat{b}_0$ is the annihilation operator of bosons. The Green's function is obtained that 

\begin{equation}
\label{eq:413}
G(\mathbf{k},\omega)=\frac{1}{i\omega-E_\mathbf{k}+i\Gamma}
\end{equation}

where $\Gamma$ is the decay rate of the electrons evaluated in the Ref. \cite{Yang}. For convenience of calculations, the decay rate $\Gamma$ is neglected. This may result in a slight overestimation of the superconducting gap $\Delta$ and disappearance of the phase factor. Also with the perturbation theory, the expectation value of $\hat{b}_0$ is

\begin{equation}
\label{eq:414}
<\hat{b}_0>=\frac{\Delta_s}{\xi_0-\mu_b}
\end{equation}

where $\xi_0$ is the ground state energy of bosons and $\mu_b$ is the chemical potential of bosons. In the scenario of the formation and decay of electron pairs, the conservation of energy demands that $\omega+E_\mathbf{k}=\xi_\mathbf{t}$ where$E_\mathbf{k}=\sqrt{\epsilon^2_\mathbf{k}+\Delta^2}$ is the dispersion relation of bare fermions and $\epsilon_\mathbf{k}=\frac{k^2}{2m}-\mu_f$. $m$ is the effective mass of electrons and $\Delta$ is the superconducting gap induced by the main superconducting mechanism. Notice that the conservation of energy results in the relation that $\xi_0-\mu_b\sim \Delta$, which can facilitate the estimation of the interaction potential $V_1$ and, as well, the superconducting gap $\Delta$. Substituting Eq. \ref{eq:412}, Eq. \ref{eq:413} and Eq. \ref{eq:414} into Eq. \ref{eq:411}, the energy gap $\Delta_s$ is cancelled and the Eq. \ref{eq:411} becomes

\begin{equation}
\label{eq:dt}
\Delta=\sum_\mathbf{k}\frac{|V_1(q=0)|^2}{2 n_0 E_{\mathbf{k}}}\tanh\frac{\beta E_\mathbf{k}}{2}.
\end{equation}

\subsection{results and discussions}
\begin{figure}
  \includegraphics[width=250pt]{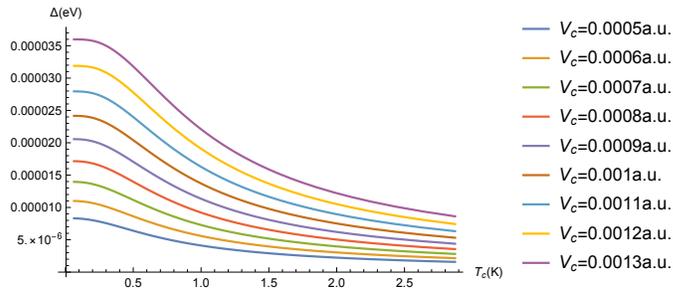}
  \caption{The relation of $\Delta$ versus $T_c$ is plotted with different values of $V_c$}
  \label{dt}
\end{figure}

By calculating and substituting the value of $\Delta$ into the right-hand side of Eq. \ref{eq:dt}, the numerical results of the superconducting gap $\Delta$ can be obtained and plotted in Fig. \ref{dt}. It is shown that the superconducting gap reduces and persists into the "normal" state of the STO/LAO system. This is in qualitative agreement with the experiments, while in experiments the superconducting gap decreases more sharply than that in the theoretical fitting. The difference may come from the presence of other interactions, such as impurity scattering, electron-electron interactions, etc. The inaccuracy from the perturbation theory and approximations also contributes another factor of the difference. In contrast with BCS theory and Eliashberg theory, the formalism can produce the superconducting gap above $T_c$ and provide another method to calculate $\Delta-T_c$ relations. This may be helpful for constructing theories for high-temperature superconductors.

Notice that the superconducting gap $\Delta$ should not be considered to be caused by the interaction potential $V_1$, since when $V_1$ vanishes, the relation of Eq. \ref{eq:411} does not hold either. In addition, if $V_c$ is very large, the perturbation theory cannot be applied in the case. Physically the two superconducting microscopic origins of $\Delta$ and $V_1$ are different. In Ref. \cite{josephson} both the microscopic origins are considered to lead to the same value of superconducting gap. This results in the misleading conclusion that both caused superconducting gaps are $\sim 1{\rm neV}$. On the contrary, if the two superconducting mechanisms are treated separately in Ref. \cite{josephson}, the main superconducting gap $\Delta$
will be around $40 {\rm \mu eV}$ and the other induced by the interaction potential $V_1$ is about $0.2 {\rm \mu eV}$. This is more feasible than the previous conclusions.

\section{conclusion}
In the present paper, one possible microscopic origin of the lifetime effects at the STO/LAO interface is proposed. Phonon-electron interactions is supposed to be responsible for the formation and decay of electron pairs, while it does not exclude the probability that other interactions may participate in the process, e.g., electron-electron interactions, impurity scattering, etc. It is discovered that if the interaction kernel is zero at Fermi surface and finite in other regions, it may well produced the even/odd free energy difference $\delta F_{e/o}$ and the effective excitation number $N_{\rm eff}$, and quantitatively describe the even/odd effects at the STO/LAO interface accurately. The interaction kernel induced by the phonon-electron interactions exactly fall into the categories of this kind of interaction kernels and is the candidate to explain microscopic origin of the lifetime effects at the STO/LAO interface. Furthermore, within the framework of the perturbation theory, the $\Delta-T_c$ relation is calculated. The result is in qualitative agreement with the experiments, while in experiments the superconducting gap $\Delta$ drops more sharply in the theory, which may be caused by the inaccuracy of perturbation and approximations. This over-simplified model may also be one of the factors. Further theoretical and experimental investigations are needed.

Beside that, the formalism presented by the paper describes the superconducting gap $\Delta$ above $T_c$ in contrast with the BCS and Eliashberg theory. This offers one novel calculation method for superconductors with the lifetime effects and superconducting gap above $T_c$ and may be helpful in analyzing high-temperature superconductors.

\bibliography{draftBib}
\bibliographystyle{unsrt}

\end{document}